\documentclass[prd,eqsecnum]{revtex4}
\usepackage{graphicx}
\usepackage{dcolumn}
\usepackage{bm}
\newcommand{\be}{\begin{equation}}
\newcommand{\ee}{\end{equation}}
\newcommand{\bse}{\begin{subequations}}
\newcommand{\ese}{\end{subequations}}
\newcommand{\bary}{\begin{eqnarray}}
\newcommand{\eary}{\end{eqnarray}}

\begin{document}

\preprint{ICN/000-03-HEP}

\title{Neutrino oscillation in Magnetized Gamma-Ray Burst Fireball}

\author{Sarira Sahu$^1$, Nissim Fraija$^1$ and Yong-Yeon Keum$^{2,3}$ }
 \affiliation{$^1$Instituto de Ciencias Nucleares, 
Universidad Nacional Aut\'onoma de M\'exico,\\ 
Circuito Exterior, C. U., A. Postal 70-543, 04510 M\'exico DF, M\'exico\\
$^{2}$Asia Pacific Center for Theoretical Physics
POSTECH, \\ San-31, Hyoja-Dong, NamKu, Pohang, Gyeongbuk 790-784 Korea,\\
$^{3}$ Department of Physics and  BK21 Initiative for Global Leaders in Physics,  
Korea University, Seoul 136-701, Korea}

\begin{abstract}

Neutrinos of energy about 5-20 MeV are produced due to the stellar collapse or
merger events that trigger the Gamma-Ray Burst. Also low energy MeV neutrinos
are produced within the fireball due to electron positron annihilation and
nucleonic bremsstrahlung. Many of these neutrinos will propagate through the
dense and relativistic magnetized plasma of the fireball. We have studied the
possibility of resonant oscillation of $\nu_e\leftrightarrow \nu_{\mu,\tau}$
by taking into account the neutrino oscillation parameters from SNO,
SuperKamiokande and Liquid Scintillator 
Detector. Using the resonance condition we have calculated the resonance length
for these neutrinos and also the fireball observables like lepton asymmetry
and the baryon load are estimated based on the assumed fireball radius of 100 Km.

\end{abstract}

\maketitle
 
\section{Introduction}

Gamma-Ray Bursts (GRBs) are flashes of non-thermal bursts of low energy 
($\sim$ 100 KeV-1 MeV) photons and release about $10^{51}$-$10^{53}$ erg in a 
few seconds making them the most luminous object in the universe after the Big
Bang\cite{Piran:1999kx,Zhang:2007nka}.  They have cosmological 
origin\cite{Piran:1999kx,Zhang:2007nka,Zhang:2003uk,Piran:1999bk} and fall   
into two classes: short-hard bursts ($\le 2~s$) and long-soft
bursts. It is now widely accepted that long duration bursts are 
produced due to the core collapse of massive stars the so called hypernovae
\cite{Piran:1999kx,Meszaros:1999fr,Ruffert:1998qg}.
The origin of short-duration bursts are still a mystery, but recently there 
has been tremendous progress  due to accurate localization of many short
bursts by the Swift\cite{Gehrels:2005qk,Barthelmy:2005bx} and
HETE-2\cite{Villasenor:2005xj} 
satellites and the observations seem to support the coalescing of
compact binaries as the progenitor for the short-hard bursts.
Recently millisecond magnetars have been considered as possible candidates as
the progenitor for the short-hard bursts\cite{Usov:1992zd,Uzdensky:2007uf}.

Irrespective of the nature of the progenitor, it is believed that, gamma-ray
emission arises from the collision of different internal shocks (shells) due 
to relativistic outflow from the source. A class 
of models called {\it fireball model} seems to explain the temporal structure
of the bursts and the non-thermal nature of their 
spectra\cite{Goodman:1986az,Piran:1999kx,Zhang:2007nka,Zhang:2003uk,Waxman:2003vh}. A major
setback of this approach is its inability to explain the late activity of the
central engine\cite{Zhang:2005fa,Burrows:2005ww}.  

In the standard fireball scenario, a radiation dominated plasma is formed in a
compact region with a size $c\delta t\sim 100$-$1000$
km\cite{Piran:1999kx,Waxman:2003vh}. This creates an opaque $\gamma-e^{\pm}$
fireball 
due to the process $\gamma+\gamma\rightarrow e^+ + e^-$. The average optical
depth of this process is $\tau_{\gamma\gamma}\simeq 10^{13}$. Because of this huge
optical depth, photons can not escape freely and even if there are no pairs to 
begin with, they will form very rapidly and will Compton scatter lower energy
photons. In the fireball the $\gamma$ and $e^{\pm}$ pairs will thermalize
with a temperature of about 3-10 MeV. The fireball  expands relativistically
with a large Lorentz factor and cools adiabatically due to the expansion. The
radiation emerges freely to the inter stellar medium (ISM), when the optical
depth is $\tau_{\gamma\gamma}\simeq 1$. In addition to $\gamma$, $e^{\pm}$
pairs, fireball may also contain some baryons, both from the progenitor and
the surrounding medium and the electrons associated with the matter (baryons)
can increase the opacity, hence delaying the process of emission of radiation.

As discussed above, core collapse of massive stars, merger of binary
compact objects (neutron star-neutron star, neutron star-black hole) and
millisecond magnetars as possible progenitors of the long and short GRBs
respectively. The process of collapse in all these scenarios are  similar to
the one that takes place in supernovae of type II and neutrinos of energy 5-20
MeV are produced. Also due to nucleonic bremsstrahlung $NN\rightarrow
NN\nu{\bar \nu}$ as well as electron positron annihilation $e^{+}e^{-}
\rightarrow Z \rightarrow \nu{\bar \nu}$, neutrinos of all the 
three flavors  can be produced during the merger
process\cite{Raffelt:2001kv}. Fractions of 
these neutrinos will be able to propagate through the fireball formed far away
from the  central engine. Within the fireball the inverse beta
decay of proton $p + e^{-} \rightarrow n + \nu_{e}$ 
will also produce MeV neutrinos which then propagate through
it. From the accretion disc neutrinos of similar energy are radiated as
discussed in the ref.\cite{Ruffert:1998qg} and fractions of these neutrinos 
may also pass through
the fireball if the accreting materials survive for longer period. 
In the fireball picture, a substantial fraction of the baryon kinetic energy
is transferred to a non-thermal population of electrons through Fermi
acceleration at the shock and these accelerated electrons will cool through
synchrotron emission and/or inverse Compton scattering to produce observed
emission in prompt and afterglow phase. The synchrotron emission from
relativistic electrons take place either in a globally ordered magnetic field
which was probably carried from the central engine or in random magnetic
fields generated in the shock dissipation region. But it is difficult to
estimate the strength of the magnetic field from the first principle. 
However polarization information of the GRBs, if retrieved, would give valuable
information regarding the magnetic field and the nature of the central engine.

The neutrino properties get modified when it propagates in a medium. Even a
massless neutrino acquires an effective mass and an effective potential in the
medium. The resonant conversion of neutrino from one flavor to another due to
the medium effect is important for solar neutrinos which is well known as the
MSW effect. Similarly the propagation of neutrino in the early universe hot
plasma\cite{Enqvist:1990ad}, supernova medium\cite{Sahu:1998jh} and in the
GRB fireball\cite{Sahu:2005zh} can have also 
many important implications in their respective physics.
The magnetic field is intrinsically entangled with the matter in all the above
scenarios. Although neutrino can not couple directly to the magnetic field,
its effect can be felt through coupling to charge particles in the background.  
Neutrino propagation in a neutron star in the presence of a magnetic field and
also in the magnetized plasma of the early universe has been studied
extensively. But to the best of our knowledge, there exist no work on the
propagation of neutrino in a magnetized fireball plasma and we believe that the
combine effect of matter and magnetic field will give interesting effect.
In this context, we have studied the propagation of low energy MeV
neutrinos in the magnetized plasma of the GRB fireball.

The paper is organized as follows: We derive the effective potential for a
neutrino in the presence of a weakly magnetized electron positron plasma in
sec. 2. In sec. 3, the effective potential for extremely strong field limit
is discussed. We discuss about the physics of GRB in sec. 4 and sec. 5 is
devoted to the oscillation of neutrinos in GRB environment by taking into
account the results from SNO, SuperKamiokande and LSND. A brief conclusion is
given in sec. 6.

\section{Neutrino Potential}
The neutrino propagation in a heat bath has been studied
extensively\cite{Enqvist:1990ad,Garcia:2007ij}. 
Using the finite temperature field theory method and considering the effect of
magnetic field through Schwinger's  propertime
method, the effective potential of a propagating neutrino is derived in a
magnetized medium\cite{Bravo Garcia:2007uc}, which can be given by, 
\be
V_{\rm eff} = b -  c \cos \phi
+ (a_\parallel - a_\perp) {|{\bf k}|} \sin^2 \phi \,
\label{poten1}
\ee
where $a$, $b$ and $c$ are the Lorentz scalars. For an electron neutrino
propagating in the above medium, the scalar functions are given by
\bary
a_\parallel&=&-\frac{g^2 e B}{M_W^4}\int_0^\infty
\frac{dp_3}{(2\pi)^2}\sum_{n=0}^{\infty} (2-\delta_{n,0})\frac{m^2}{E_{e,n}}(f_{e,n}+\bar{f}_{e,n})\cr
&&+\frac{g^2}{4M_W^4}\biggl(k_3(N^0_e-\bar{N}^0_e)+k_0(N_e-\bar{N}_e) \biggr),
\eary
\bary
a_\perp&=&-\frac{g^2 e B}{M_W^4}\int_0^\infty
\frac{dp_3}{(2\pi)^2}\sum_{n=0}^{\infty}(2-\delta_{n,0})\biggl(\frac{2neB}{2E_{e,n}}
+\frac{m^2}{E_{e,n}}\biggr)(f_{e,n}+\bar{f}_{e,n})\nonumber\\ 
&&+\frac{g^2}{4M_W^4}\biggl(k_3(N^0_e-\bar{N}^0_e)+k_0(N_e-\bar{N}_e) \biggr),
\eary
\bary
b
&=&\frac{g^2}{4M_W^2}(N_e-{\bar N}_e)(1+c_V)+\frac{g^2eB}{4M_W^4}
(N_e^0-{\bar N}_e^0)\nonumber\\ 
&&-\frac{eBg^2}{M^4_W}\int_0^\infty\frac{dp_3}{(2\pi)^2}\sum_{n=0}^{\infty}
(2-\delta_{n,0})
\bigl[\frac{k_3}{E_{e,n}}(p_3^2+\frac{m^2}{2})\delta_{n,0}+
 E_{\nu_e}E_{e,n}\bigr](f_{e,n}+{\bar{f}}_{e,n}), 
\label{bcompl}\\
c&=&\frac{g^2}{4 M_W^2}(N_e^0-{\bar N}_e^0)(1-c_A)+\frac{g^2eB}{4M_W^4}
(N_e-{\bar N}_e)\nonumber\\
&&-\frac{eBg^2}{2M_W^4}\int_0^\infty \frac{dp_3}{2\pi^2}\sum_{n=0}^{\infty}
(2-\delta_{n,0})
\biggl[E_{\nu_e}(E_{e,n}-\frac{m^2}{E_{e,n}})\delta_{n,0}
+\frac{k_3p_3^2}{E_{e,n}}\biggr](f_{e,n}+{\bar{f}}_{e,n}).
\label{ccompl}
\eary
In the magnetic field, the electron energy is given by
\be
E^2_{e,n}=(p^2_3+ m^2+2ne B).
\ee
For electrons in the background we have $c_V=-\frac{1}{2}+2 \sin^2\theta_W$,
$c_A=\frac{1}{2}$, $m$ is the electron mass and $B$ is the constant background
magnetic field. 
In Eq.\, (\ref{poten1}), $\phi$ is the angle between the
neutrino momentum and the direction of the magnetic field (${\bf k\cdot B}$). We
will 
be considering the forward moving neutrinos (or moving along the magnetic field)
and in rest of the paper consider $\phi\simeq 0$. Also for massless neutrino
we assume $k_0=k_3=E_{\nu}$. For neutrinos propagating in the
forward direction the last term in the Eq.(\ref{poten1}) vanishes. 
Also for the strong magnetic field case, when only the lowest
Landau level ($n=0$) is populated the term $(a_\parallel - a_\perp)$ vanishes.   
Thirdly even if the neutrinos are not propagating in the forward direction
but the magnetic field is weak, then the term $(a_\parallel - a_\perp)$ is
very small. So we neglect its contribution in the rest of the paper and with
this the effective potential can be given by
\be
V_{\rm eff} = b -  c \cos \phi.
\ee
We shall assume that the magnetic field is weak ($B\ll m^2/e=B_c$) in the
electron-positron plasma where the test neutrino is propagating. The
electron density in a magnetized plasma is given by
\bary
N_e &=&\frac{2eB}{4\pi^2}\sum_{n=0}^\infty (2 - \delta_{n,0}) \int_0^\infty dp_3
f_{e,n}\nonumber\\
&=&\frac{2eB}{4\pi^2}\bigl[2\sum_{n=0}^\infty \int_0^\infty dp_3 f_{e,n}-\int_0^\infty dp_3 f_{e,0} \bigr],
\eary
where we can further define 
\be
N_e^0=\frac{2eB}{4\pi^2}\int_0^\infty dp_3 f_{e,0}.
\ee
We also assume that
the chemical potential ($\mu$) of the electrons and positrons are much smaller
than their energy i.e. $\mu\ll E_e$. In this case the fermion distribution
function can be written as a sum and is given by
\be
f(E_e)=\frac{1}{e^{\beta(E_e-\mu)}+1}\simeq \sum^{\infty}_{l=0} (-1)^l
e^{-\beta(E_e-\mu)(l+1)}.
\ee
Using the above distribution function, the electron number density in the weak
field limit is
\be
N_e=\frac{m^3}{2\pi^2} \sum_{l=0}^{\infty} (-1)^l e^{\alpha}
\left[
\frac{2}{\sigma} K_2(\sigma)-\frac{B}{B_c} K_1(\sigma)
\right],
\ee
and
\be
N^0_e=\frac{1}{2\pi^2}\frac{B}{B_c} m^3 \sum_{l=0}^{\infty} (-1)^l e^{\alpha}
 K_1(\sigma).
\label{ne0}
\ee
where we have defined
\bary
\alpha &=& \beta\mu (l+1),\nonumber\\
\sigma &=& \beta m (l+1),
\eary
and $K_i$ is the modified Bessel function of integral order $i$. With the help
of above, for an electron neutrino propagating in the medium, the Lorentz
scalars $b$ and $c$ are expressed as
\bary
b &=& b_0-\frac{4\sqrt{2}}{\pi^2} G_F\left (\frac{m}{M_W}\right )^2 m^2
E_{\nu_e}\sum_{l=0}^{\infty} (-1)^l \cosh\alpha
\left [
\left (\frac{3}{\sigma^2} - \frac{1}{4}\frac{B}{B_c}\right ) K_0(\sigma)
+\left (1+\frac{6}{\sigma^2} \right ) \frac{K_1(\sigma)}{\sigma}
\right ], \nonumber\\
\label{defnb}
c &=& c_0-\frac{4\sqrt{2}}{\pi^2} G_F\left (\frac{m}{M_W}\right )^2 m^2
E_{\nu_e}\sum_{l=0}^{\infty} (-1)^l  \cosh\alpha
\frac{1}{\sigma^2}
\left (K_0(\sigma)+\frac{2}{\sigma}K_1(\sigma) \right ),
\label{defnc}
\eary
where
\bary
b_0 &=& \sqrt{2} G_F\left [(N_e-{\bar N}_e) (1+c_V) + \frac{B}{B_c} 
\left (\frac{m}{M_W}\right )^2 (N_e^0-{\bar N}_e^0)
\right ] ,\nonumber\\
\label{defnb0}
c_0 &=& \sqrt{2} G_F\left [(N^0_e-{\bar N}^0_e) (1-c_A) + \frac{B}{B_c} 
\left (\frac{m}{M_W}\right )^2 (N_e-{\bar N}_e)
\right ].
\label{defnc0} 
\eary
For muon and tau neutrinos, only the neutral current interaction will
contribute. So for $\nu_{\mu,\tau}$ only $c_V$ and $c_A$ terms will
contribute. For anti-neutrino,  $(N^0_e-{\bar N}^0_e)$ will be replaced by
$-(N^0_e-{\bar N}^0_e)$ and similarly $(N_e-{\bar N}_e)$ by  
$-(N_e-{\bar N}_e)$. For our convenience we can also define
\be
N^0_e-{\bar N}^0_e=\frac{m^3}{\pi^2}\frac{B}{B_c} \sum_{l=0}^{\infty}
(-1)^l \sinh{\alpha}\, K_1(\sigma) =\frac{m^3}{\pi^2}\Phi_1, 
\label{numden0}
\ee
and 
\label{sfen}
\be
N_e-{\bar N}_e=\frac{m^3}{\pi^2} \sum_{l=0}^{\infty} (-1)^l
\sinh{\alpha}\, 
\left[
\frac{2}{\sigma} K_2(\sigma)-\frac{B}{B_c} K_1(\sigma)
\right]=\frac{m^3}{\pi^2} \Phi_2.
\label{numdenb}
\ee
 In the weak field limit, the effect of magnetic field is very small in
 $N_e-{\bar N}_e$ and it is important when $B\gg B_c$. Due to the weak field
 limit, the magnetic field contribution is very much suppressed  which is shown
 in Eq.\,(\ref{numdenb}).  But in strong
 field limit, which is described in the next section, the number density is
 proportional to the magnetic field Eq.\,(\ref{numden0}).

\section{Very Strong field limit}

For very strong magnetic field, only the lowest Landau Level (LL)
$n=0$ will contribute and in this case the energy of the particle is
independent of the magnetic field and can be given by 
\be
E^2=(p^2_3+m^2),
\ee
and the number density of electrons is given by Eq.\,(\ref{ne0}). Defining the 
particle asymmetry in the background as
\be
L_i=\frac{(N_i-{\bar N}_i)}{N_{\gamma}},
\ee
and also we have define $L^0_i$ when the particles are in LL. 
where $N_{\gamma}=2/\pi^2 \xi(3) T^3$ is the photon number density, we can
express
\bary
b_0&=&\sqrt{2} G_F N_{\gamma} 
\left [L_e (1+c_V)+ \frac{B}{B_c} \left (\frac{m}{M_W}\right )^2 L^0_e
\right ],\nonumber\\
c_0&=&\sqrt{2} G_F N_{\gamma} \left [L^0_e (1-c_A)+ \frac{B}{B_c} \left
    (\frac{m}{M_W}\right )^2 L_e \right ],
\eary
and the potential can be written as
\bary
V &=& \sqrt{2} G_F N_{\gamma} L_e^0 \left [ 1+c_V+\frac{B}{B_c} \left (
    \frac{m}{M_W} \right )^2 -\left( 1-c_A+\frac{B}{B_c} 
 \left ( \frac{m}{M_W} \right )^2 \right ) \cos\phi\right ]\nonumber\\
&&-\frac{2\sqrt{2}}{\pi^2} G_F \frac{B}{B_c}  
\left ( \frac{m}{M_W} \right )^2 m^2 E_{\nu} 
\sum_{l=0}^{\infty} (-1)^l  \cosh\alpha
\left [
\left (\frac{3}{2} K_0(\sigma) +
  \frac{2}{\sigma} K_1(\sigma) \right )-\frac{K_1(\sigma)}{\sigma}\cos\phi
\right ].
\eary
For forward moving neutrinos, the potential is simplified to
\bary
V &=& \sqrt{2} G_F N_{\gamma} L_e^0 ( c_V+c_A)\nonumber\\
&& -\frac{2\sqrt{2}}{\pi^2} G_F \frac{B}{B_c}  
\left ( \frac{m}{M_W} \right )^2 m^2 E_{\nu} 
\sum_{l=0}^{\infty} (-1)^l  \cosh\alpha
\left (\frac{3}{2} K_0(\sigma) +\frac{K_1(\sigma)}{\sigma}\right ).
\eary
This is the potential for $\nu_e$ propagating in the strongly magnetized 
$e^-e^+$ plasma ($c_V$ and $c_A$ are already defined for electron background),
whereas for $\nu_{\mu}$ 
and $\nu_{\tau}$ the last term is absent which is order of magnitude
suppressed. So for a system where lepton asymmetry is non-zero one can neglect
the second term. In this situation the active-active neutrino oscillation is
very much suppressed due to the cancellation of the leading order term. 
The magnetars or anomalous X-ray pulsars (AXPs) are believed to 
have magnetic field much above the critical field $B >> m^2/e$ and 
probably one can use the above potential to study the neutrino
propagation in their magnetized environment.

\section{GRB physics}

 A fireball is formed due to the sudden release of copious amount of
 $\gamma$-rays into a compact region with a size $c\delta t\sim 100-1000$ km
 by creating an opaque $\gamma-e^-e^+$ plasma. In the fireball the $\gamma$s
 and pair plasma will thermalize with a temperature of about 3-10
 MeV. Afterward the fireball will expand relativistically under its own
 pressure and cools
 adiabatically\cite{Piran:1999kx,Zhang:2003uk,Meszaros:1999fr} . When the
 optical depth of photon is of 
 order unity, the radiation emerges freely to the intergalactic medium. As
 stated above, we shall consider the fireball temperature in the range 3-10
 MeV for our analysis.

Baryon load in the fireball is an outstanding issue. The fireball contains
baryons both from the progenitor and the 
surrounding medium. The electrons associated with the matter (baryons) can 
increase the opacity, hence delaying the process of radiation emission.
The baryons can be accelerated along with the fireball and convert part of the
radiation energy into bulk kinetic energy.
So the dynamics of the fireball crucially depends on the baryon content of it.
But the baryon load of the fireball has to be
low ($10^{-8}M_{\odot}-10^{-5}M_{\odot}$) otherwise it will be Newtonian and
there will be no GRB\cite{Piran:1999kx,Waxman:2003vh}.

Here we consider a CP-asymmetric $\gamma$ and $e^-e^+$ fireball, 
where the excess of electrons come from the electrons associated
with the baryons within the fireball. We have shown earlier for $B=0$
case that for the active-active neutrino oscillation, the potential is
independent of the baryonic contribution. However for active-sterile neutrino
oscillation the potential does depend on the baryonic
contribution\cite{Sahu:2005zh}.  

The problem of magnetic field in the GRBs is
outstanding\cite{Zhang:2003uk}. There 
is no way to get the magnetic field information directly from the
fireball. It is strongly believed in the GRB community that the
$\gamma$-rays which we detect  are mostly due to the synchrotron
radiation of charged particles in the magnetic field although the strength of
it is still unknown. But the field strength will be smaller than $B_c$,
because even if the central engine is having very strong magnetic field, the
magnetic field will decay as $\sim r^{-2}$ when the jet moves away from the
central engine making it weak. 
Here we shall take the weak field approximation $B\ll B_c$ and study
the oscillation of neutrinos in the fireball environment.

In a stellar collapse or merger of compact binaries 5-20 MeV neutrinos are
produced that trigger the burst. Due to nucleonic bremsstrahlung and
annihilation of $e^-e^+$, neutrinos of all kinds are produced which has a low
flux compared to the previous process. Also due to inverse beta decay process
MeV neutrinos can be produced. Normally the 5-20 MeV neutrinos produced due to
collapse or merger of compact binaries will go away before the fireball is
formed. If the fireball is fed continuously with the late time ejecta
powered by neutrinos from the accretion torus then some of these MeV neutrinos
will propagate through the fireball which we have discussed in the
introduction. So due to above neutrino production 
mechanisms whether external or internal to the fireball some of these  
neutrinos will propagate through the fireball and the fireball plasma being in
an extreme condition may affect the propagation of these neutrinos through it.

  The GRBs are also sources of very high energy neutrinos and gammas which are
  produced during different stages of its dynamical evolution. Bahcall and
  Meszaros\cite{Bahcall:2000sa} have shown that due to dynamical decoupling of
  neutron from the 
  rest of the fireball plasma, inelastic collision of protons and neutrons
  will produce 5-10 GeV neutrinos and they estimate about 7 events per year in
  a $km^3$ detector (for redshift $z\simeq 1$). But production of these
  neutrinos crucially depends on the neutron content of the fireball.

Also two different mechanisms are discussed by Meszaros and
Rees\cite{Meszaros:2000fs}  for the production of 2-25 GeV neutrinos. In the
first mechanism they show that if internal shocks occur below
   the radiation photosphere, rapid diffusion of neutrons in both parallel and
   transverse to the radial direction occurs and inelastic collision with the
   protons can give rise to pions and subsequently neutrinos of energy about 2
   GeV. In the second mechanism, neutrons diffuse transversely from a slower
   outflow into a fast jet, at a height where the transverse inelastic optical
   depth of the jet is close to unity. This mechanism can produce neutrino
   energy of order 25 GeV. The $km^3$ detectors with sufficiently dense
   phototubes can be able to detect about 3-15 events per year for $z\ge 1$.

The high energy gamma radiation can also be observed from GRB by acceleration
of high energy protons in the magnetic field and at the same time accelerated
high energy protons 
can also produce very high energy neutrinos\cite{Waxman:1997ti,Abbasi:2009kq}
and gamma rays\cite{Morris:2006rr,Galli:2008uz,Abdo:2009zz,Corsi:2009ib} due
to photo pion production as well as proton-proton collisions. All these
photons and neutrinos are in principle observables with the present day
detectors.

Here for simplicity we assume that the fireball is charge neutral $L_e=L_p$
and spherical
with an initial radius $R\simeq (100-1000)$ km and it has equal number of
protons and neutrons. Then the baryon load in the fireball can be given by
\bary
M_b &\simeq & \frac{16}{3\pi} \xi(3) L_e T^3 R^3 m_p\nonumber\\
    &\simeq & 2.23\times 10^{-4} L_e T^3_{MeV} R^3_7 M_{\odot}.
\eary
where $T_{MeV}$ is the fireball temperature expressed in MeV and $R_7$ is in
units of $10^7$ cm and $m_p$ is the proton mass. For ultra relativistic
expansion of the fireball, we assume the baryon load in it to be in the range 
$10^{-8}M_{\odot}-10^{-5}M_{\odot}$ which corresponds to lepton asymmetry
in the range $8.1\times 10^{-4} R^{-3}_7\le L_e \le 8.1\times 10^{-1} R^{-3}_7$. 

\section{neutrino oscillation}

Here we consider the neutrino oscillation process $\nu_e\leftrightarrow
\nu_{\mu, \tau}$.
The evolution equation for the propagation of neutrinos in the above medium
is given by
\be
i
{\pmatrix {\dot{\nu}_{e} \cr \dot{\nu}_{\mu}\cr}}
={\pmatrix
{V-\Delta \cos 2\theta & \frac{\Delta}{2}\sin 2\theta \cr
\frac{\Delta}{2}\sin 2\theta  & 0\cr}}
{\pmatrix
{\nu_{e} \cr \nu_{\mu}\cr}},
\ee
where $\Delta=\delta m^2/2E_{\nu}$, $V$ is the potential difference
between $V_{\nu_e}$ and $V_{\nu_{\mu}}$, (i. e. $V=V_{\nu_e}-V_{\nu_{\mu}}$),
    $E_{\nu}$ is the neutrino energy and $\theta$ is the neutrino 
mixing angle. The conversion probability for the above process at a given time
$t$ is given by
\be
P_{\nu_e\rightarrow {\nu_{\mu}{(\nu_\tau)}}}(t) = 
\frac{\Delta^2 \sin^2 2\theta}{\omega^2}\sin^2\left (\frac{\omega t}{2}\right
),
\label{prob}
\ee
with
\be
\omega=\sqrt{(V-\Delta \cos 2\theta)^2+\Delta^2 \sin^2
    2\theta}.
\ee
The potential for the above oscillation process is
\bary
V&=&\sqrt{2} G_F \frac{m^3}{\pi^2} \left [  
\Phi_1 -\Phi_2 \cos\phi + \frac{B}{B_c}\left (\frac{m}{M_W}\right )^2
(\Phi_2-\Phi_1 \cos\phi)\right.\nonumber\\
&&\left.-\frac{4}{\pi^2} 
\left (\frac{m}{M_W}\right)^2 \frac{E_{\nu_e}}{m}
(\Phi_3-\Phi_4 \cos\phi) 
\right ],
\label{poten2}
\eary
where we have defined
\bary
\Phi_3&=&\sum_{l=0}^{\infty} (-1)^l
\cosh{\alpha}\, \left [\left (\frac{3}{\sigma^2}-\frac14
  \frac{B}{B_c}\right )K_0(\sigma)+\left (1+\frac{6}{\sigma^2}\right )
\frac{K_1(\sigma)}{\sigma}\right ],\nonumber\\  
\Phi_4&=& \sum_{l=0}^{\infty} (-1)^l
\cosh{\alpha}\,  \frac{1}{\sigma^2}
\left [K_0(\sigma)+\frac{2}{\sigma}K_1(\sigma)\right ], 
\label{phies}
\eary
and $\Phi_1$ and $\Phi_2$ are defined in Eqs.\,(\ref{numden0}) and
(\ref{numdenb}). For $\phi\simeq 0$ and weak field limit the potential can be
written as 
\be
V \simeq \sqrt{2} G_F \frac{m^3}{\pi^2} \left [  
\Phi_1 -\Phi_2 -\frac{4}{\pi^2} 
\left (\frac{m}{M_W}\right)^2 \frac{E_{\nu_e}}{m}
(\Phi_3-\Phi_4) 
\right ].
\ee
For anti-neutrinos the functions $\Phi_1$ and $\Phi_2$ will change signs.
The oscillation length for the neutrino is given by
\be
L_{osc}=\frac{L_v}{\sqrt{\cos^2 2\theta (1-\frac{V}{\Delta \cos 2\theta}
    )^2+\sin^2 2\theta}},
\label{osclength}
\ee
where $L_v=2\pi/\Delta$ is the vacuum oscillation length. 
For resonance to occur, we should have $V>0$ and   
\be
V=\Delta \cos 2\theta.
\label{reso}
\ee
The resonance length can be given by
\be
L_{res}=\frac{L_v}{\sin 2\theta}.
\ee
The positivity of the potential implies that the chemical potential  $\mu$ of
the background electrons and positions should not be zero, so that the
difference of the number densities of the particles and anti-particles as
shown in Eqs. (\ref{numden0}) and  (\ref{numdenb}) will be non vanishing. Also
$\mu$ should not be very  small, otherwise the potential will be negative. The
resonance condition is 
\be
\Phi_1 -\Phi_2 -3.196\times 10^{-11} E_{MeV}
(\Phi_3-\Phi_4) =2.26 \frac{\tilde {\delta m^2}}{E_{MeV}} \cos 2\theta,
\label{rescond}
\ee
where ${\tilde {\delta m^2}}$ is expressed in units of $eV^2$ and the neutrino
energy $E_{\nu}$ in units of MeV as $E_{MeV}$.
The left hand side depends on $\mu$, temperature $T$ of the plasma and the
neutrino energy. On the other hand the right hand side depends only on the
neutrino energy (for a given set of neutrino mass square difference and the
mixing angle). Let us emphasize that the resonance condition for $B=0$ and
$B\neq 0$ are different. 
In the $B=0$ case for resonance condition to satisfy,
first the lepton asymmetry has to satisfy the necessary condition $L_e >
6.14\times 10^{-9} T^2_{MeV}$\cite{Sahu:2005zh}, whereas the presence of
magnetic field modifies this condition as shown in Eq. (\ref{rescond}). In the
magnetic field 
case, there is no explicit temperature dependence. Because of these
modifications the magnetized plasma result is different from the $B=0$
case. But the resonance length for both the situations are the same as the
resonance length does not depend on the magnetic field.

We have found that at resonance the function $\Phi_2$ is order
of magnitude smaller than $\Phi_1$  and $3.196\times 10^{-11} E_{MeV}
(\Phi_3-\Phi_4)$ is of same order as $\Phi_1$. At the resonance for a given
set of neutrino oscillation parameters ${\delta m^2}$ and $\sin^2
2\theta$, the resonance length only depends linearly on the neutrino
energy. So the change in background temperature or number density in the
fireball will not affect $L_{res}$. For our analysis we have taken three
different neutrino energies  $E_{\nu}=$ 5, 10 and 20 MeV and for each neutrino
energy three different fireball temperatures
$T=$ 3, 5 and 10 MeV are taken. We take into account the neutrino oscillation
parameters from solar, atmospheric (SNO and SuperKamiokande), and the Liquid
Scintillator Neutrino Detector (LSND) reactor neutrinos  to study the
resonance conditions in the fireball. The resonance  oscillation of neutrinos
can constrain the fireball parameters. For the best fit neutrino oscillation
parameter sets $\delta m^2$ and $\sin^2 2\theta$ of the above three different
state of the art experiments (SNO, Super Kamiokande and LSND), we have shown
what should be the values of $\mu$ and $T$ to satisfy the resonance condition
for different neutrino energies in the fireball plasma. Afterward these values
of $\mu$ and $T$ are used to calculate the lepton asymmetry $L_e$, baryon load
$M_b$ and the resonance length $L_{res}$ of the propagating neutrinos. 

\begin{figure}[t!] 
\vspace{0.5cm}
{\centering
\resizebox*{0.4\textwidth}{0.4\textheight}
{\includegraphics{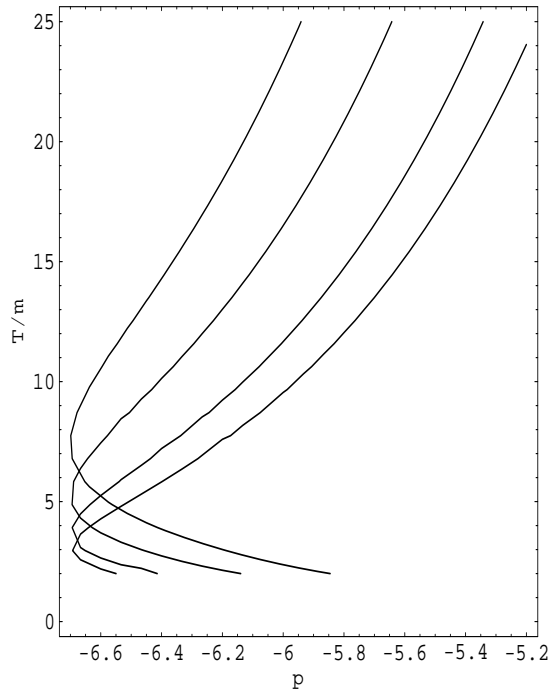}}
}
\caption{\small\sf For the best fit value of the SNO data $\delta m^2\sim
  7.1\times 10^{-5}\, eV^2$ and $\sin^2 2\theta\sim 0.69$ we have the contour
  plot for $p$ and $T/m$ (where $\mu=10^p m$) satisfying the resonance
  condition for four different neutrino energies 5, 10, 20 and 30 MeV from
  left to right respectively. Here we have taken $B/B_c=0.1$.}
\label{snores}
\end{figure}
\begin{figure}[t!] 
\vspace{0.5cm}
{\centering
\resizebox*{0.4\textwidth}{0.4\textheight}
{\includegraphics{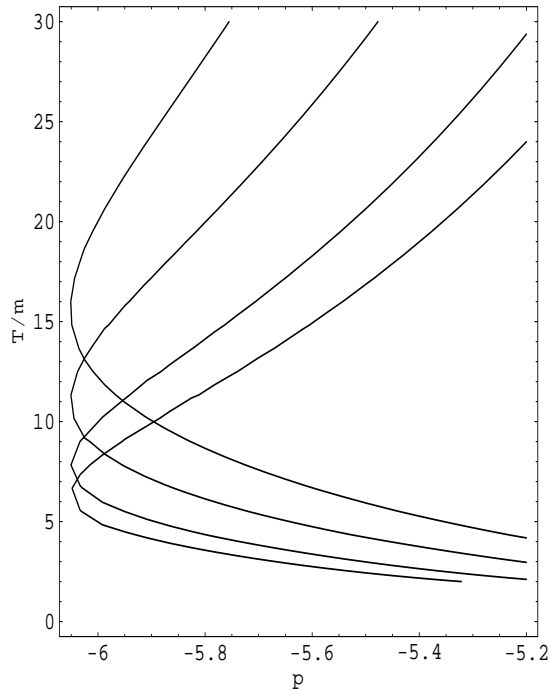}}
}
\caption{\small\sf The contour plot satisfying the resonance condition for the
  Super-Kamiokande neutrino oscillation parameters $\delta m^2\sim
  2.5\times 10^{-3}\, eV^2$ and $\sin^2 2\theta\sim 0.9$, for different $p$
  and $T/m$ are shown. The definitions of $p$ is same as in Fig.\,
  \ref{snores} and also the same magnetic field is used. The four different
  curves  from left to right are for 5, 10, 20 and 30 MeV neutrino energy
  respectively.}   
\label{skres}
\end{figure}
\begin{figure}[t!] 
\vspace{0.5cm}
{\centering
\resizebox*{0.4\textwidth}{0.4\textheight}
{\includegraphics{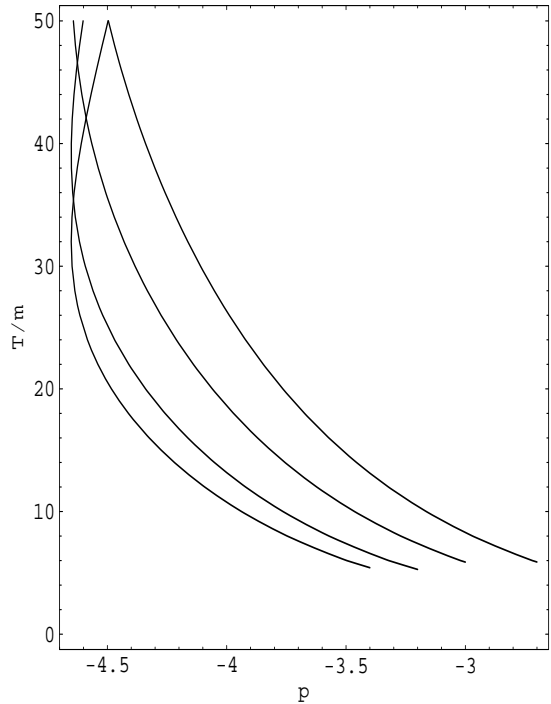}}
}
\caption{\small\sf This is for the LSND oscillation parameters $\delta m^2\sim
  0.5\, eV^2$ and $\sin^2 2\theta\sim 0.0049$ and all other parameters are the
  same as in Fig.\, \ref{snores} but the 5, 10, 20, 30 MeV curves are from
  {\bf right to left}. } 
\label{lsndres}
\end{figure}
\begin{table}
\begin{center}
\caption{SNO: The best fit values of the neutrino oscillation parameters 
$\delta m^2\sim 7.1\times 10^{-5}\, eV^2$ and $\sin^2 2\theta\sim 0.69$ from
the combined analysis of the salt phase data of SNO\cite{Ahmed:2003kj} and
KamLAND\cite{Araki:2004mb} are used in the 
resonance condition for different neutrino energies in this table. The
magnetic field used here is $B/B_c=0.1$.\\}
\renewcommand{\tabcolsep}{0.35cm}
\renewcommand{\arraystretch}{1.05}
\begin{tabular}{|c|c|c|c|c|}\hline
$E_{MeV}$ & T(MeV)         & $L_e$                  & $L_{res}$(cm)  &
$M_b(R_7^3\, M_\odot)$\\ \hline 
        & $3$              &  $3.28\times 10^{-8}$  &   &
        $2.97\times 10^{-10}$\\ 
5       & $5$              &  $4.93\times 10^{-8}$  &$2.10\times 10^{7}$   &
$9.14\times 10^{-10}$\\  
        & $10$             &  $5.07\times 10^{-8}$  &  &
        $1.13\times 10^{-8}$\\  \hline 
        & $3$              &  $4.71\times 10^{-8}$  &   &
        $2.83\times 10^{-10}$\\ 
10      & $5$              &  $5.34\times 10^{-8}$  & $4.21\times 10^{7}$ &
$1.49\times 10^{-9}$\\ 
        & $10$             &  $9.99\times 10^{-8}$  &   &
        $2.23\times 10^{-8}$\\ \hline 
        & $3$              &  $6.83\times 10^{-8}$  &  &
        $4.11\times 10^{-10}$\\ 
20      & $5$              &  $1.02\times 10^{-7}$  &$8.42\times 10^{7}$   &
$2.85\times 10^{-9}$\\ 
        & $10$             &  $1.99\times 10^{-7}$  &   &
        $4.44\times 10^{-8}$ \\ \hline 
\end{tabular}
\label{snotable}
\end{center}
\end{table}


In Fig.\ref{snores} we have shown the values of $\mu$  and $T$ which satisfy
the resonance condition for four different neutrino energies 5, 10, 20 and 30
MeV respectively by taking into account the best fit values of $\delta m^2$
and $\sin^2 2\theta$ from SNO for fireball temperature in
the range 3 to 10 MeV. In this range of temperatures it is shown that each value
of chemical potential corresponds to two different temperatures, so we can tell
that the temperature is degenerate. The increase in neutrino energy decreases
the lower temperature for a particular $\mu$ and finally the temperature
degeneracy goes away for high energy neutrinos. 
The same behavior is observed in
the temperature and the chemical potential for the best fit Super Kamiokande 
result ($\delta m^2$ and $\sin^2 2\theta$) which is plotted in
Fig.\ref{skres}. But for 
the combined LSND and KARMEN data Fig. \ref{lsndres} we have shown that there
is no temperature degeneracy observed in the range of temperature (3-10 MeV) we
consider. The degeneracy appears when the temperature goes above about 17 MeV,
which is clearly seen in  Fig. \ref{lsndres}.  Below we analyze our result for
the above three experiments separately.

{\bf SNO:} The salt phase data of SNO from solar neutrino\cite{Ahmed:2003kj}, combined with the KamLAND\cite{Araki:2004mb}
reactor anti-neutrino results constraint the neutrino oscillation parameters
and are given by $6\times 10^{-5}\, eV^2 < \delta m^2 < 10^{-4}\, eV^2$ and
$0.64 <\sin^2 2\theta < 0.96$. The best fit point for the above data is
obtained for $\delta m^2\sim 7.1\times 10^{-5} \, eV^2$ and $\sin^2 2\theta\sim
0.69$ with 99\% confidence level . We have used this best fit point for the
resonance condition for different neutrino 
energies and the observables are given in TABLE\, \ref{snotable}. For neutrino
energy 5 MeV and the fireball temperature 3 MeV, 
the lepton asymmetry is $L_e\sim 3.28\times 10^{-8}$, $L_{res}\sim 210$ km and
$M_b\sim 2.97\times 10^{-10} R^3_7M_{\odot}$. If the fireball radius is 100
Km, then the resonance length is longer than the size of the fireball and also
the baryon load is too low. The baryon load problem can be resolved by
increasing the fireball radius, but neutrino can just oscillate because still
the resonance length is quite large. 

Going from 5 MeV neutrinos to 20 MeV neutrinos and background temperature from
3 MeV to 10 MeV, we have $L_e\sim 2\times 10^{-7}$, $M_b\sim 4.44\times
10^{-8} R^3_7 M_{\odot}$ and $L_{res}\sim 842$ km. For higher energy neutrinos
the resonance length is so large that even if we increase the radius to 1000 km,
there is hardly any resonant oscillation of neutrino within the fireball.
So for neutrino oscillation parameters in the solar neutrino range (SNO) there
is hardly any resonant oscillation. 

\begin{table}
\caption{SK: The best fit values of the atmospheric neutrino oscillation
  parameters $\delta m^2\sim 2.5\times 10^{-3}\, eV^2$ and $\sin^2 2\theta\sim
  0.9$ from Super-Kaminkande Collaboration\cite{Ashie:2004mr} are used in the
resonance condition for different neutrino energies in this table. The
magnetic field we have taken here is $B/B_c=0.1$.}
\begin{center}\renewcommand{\tabcolsep}{0.35cm}
\renewcommand{\arraystretch}{1.05}
\begin{tabular}{|c|c|c|c|c|}\hline
$E_{MeV}$ & T(MeV)         & $L_e$                  & $L_{res}$(cm)  &
$M_b(R_7^3 M_\odot)$\\ \hline 
        & $3$              &  $7.04\times 10^{-7}$  &  & $4.24\times
        10^{-9}$\\ 
5       & $5$              &  $1.73\times 10^{-7}$  & $522763$  & $4.83\times
10^{-9}$\\  
        & $10$             &  $6.92\times 10^{-8}$  &   & $1.54\times
        10^{-8}$\\  \hline 
        & $3$              &  $3.81\times 10^{-7}$  &  &
        $2.30\times 10^{-9}$\\ 
10      & $5$              &  $1.30\times 10^{-7}$  & $1.05\times 10^{6}$   &
$3.62\times 10^{-9}$\\ 
        & $10$             &  $1.09\times 10^{-7}$  &   & $2.44\times
        10^{-8}$\\ \hline 
        & $3$              &  $2.38\times 10^{-7}$  & &
        $1.44\times 10^{-9}$\\ 
20      & $5$              &  $1.37\times 10^{-7}$  &$2.09\times 10^{6}$  &
$3.81\times 10^{-9}$\\ 
        & $10$             &  $2.04\times 10^{-7}$  &  & $4.54\times 10^{-8}$
        \\ \hline 
\end{tabular}
\label{sktable}
\end{center}
\end{table}

{\bf SuperKamiokande:}
The atmospheric neutrino oscillation parameters reported by the
Super-Kamiokande (SK) Collaboration\cite{Ashie:2004mr} are in the range
$1.9\times 10^{-3}\, eV^2 < 
\delta m^2 <  3.0\times 10^{-3}\, eV^2$ and $0.9\le \sin^2 2\theta \le 1.0$
with a  90\% confidence level. In this parameter space we consider the
good fit point $\delta m^2 \sim 2.5\times 10^{-3}\, eV^2$ and $\sin^2
2\theta\sim 0.9$ to study the resonance condition in the GRB fireball . The
result of our analysis is given in TABLE\, \ref{sktable}.  

For neutrino energy in the range 5 to 20 MeV and background temperature in the
range 3 to 10 MeV, there is not much variation in $L_e$ but some variation in
$L_{res}$ and in $M_b$ is observed. For neutrino energy 5
MeV and fireball temperature 5 MeV we have $L_e\sim 1.73\times
10^{-7}$ and $L_{res}\sim 5.2$ km . Similarly for neutrino energy 10 MeV and
fireball temperature 10 MeV, the $L_{res}\sim 10.5$ km, which is twice the one
for neutrino energy 5 MeV.  For 20 MeV neutrinos we obtain $L_{res}\sim 21$
km. This is 
because for a given set of neutrino oscillation parameters the resonance
length is proportional to neutrino energy and does not depends on other
factors. Also the baryon content in the fireball is proportional to $T^3$. 
So for a given neutrino energy, background with high temperature has more
baryon content than the low temperature one. The low value of $M_b$ can be
adjusted within $10^{-8}M_{\odot}$ to $10^{-5}M_{\odot}$ by adjusting the
$R_7$. Both the $L_{res}$ and $L_e$ are within the range of value we would
expect and with this $L_{res}$, before coming out of the fireball, the neutrino
can oscillate many times resonantly from one species to another.

\begin{table}
\caption{LSND: The best fit values of the neutrino oscillation parameters 
$\delta m^2\sim 0.5\, eV^2$ and $\sin^2 2\theta\sim 0.0049$ from
LSND and KARMEN\cite{Church:2002tc} are used in the resonance condition for
different neutrino 
energies.The magnetic field we have taken is $B/B_c=0.1$.} 
\begin{center}\renewcommand{\tabcolsep}{0.35cm}
\renewcommand{\arraystretch}{1.05}
\begin{tabular}{|c|c|c|c|c|}\hline
$E_{MeV}$ & T(MeV)         & $L_e$                  & $L_{res}$(cm)  &
$M_b(R_7^3 M_\odot)$\\ \hline 
        & $3$              &  $4.77\times 10^{-4}$  &  & $2.87\times
        10^{-6}$\\ 
5       & $5$              &  $9.38\times 10^{-5}$  & $35424.1$  & $2.62\times
10^{-6}$\\  
        & $10$             &  $1.23\times 10^{-5}$  &   & $2.74\times
        10^{-6}$\\  \hline 
        & $3$              &  $2.27\times 10^{-4}$  &  & $1.37\times
        10^{-6}$\\ 
10      & $5$              &  $4.77\times 10^{-5}$  & $70848.1$   &
$1.33\times 10^{-6}$\\ 
        & $10$             &  $6.36\times 10^{-6}$  &   & $1.42\times
        10^{-6}$\\ \hline 
        & $3$              &  $1.18\times 10^{-4}$  & & $7.11\times
        10^{-7}$\\ 
20      & $5$              &  $2.46\times 10^{-5}$  & $141696$ & $6.85\times
10^{-7}$\\ 
        & $10$             &  $3.28\times 10^{-6}$  &  & $7.32\times 10^{-7}$
        \\ \hline 
\end{tabular}
\label{lsndtable}
\end{center}
\end{table}

{\bf LSND:} Finally we consider the reactor neutrino data from LSND and
KARMEN\cite{Church:2002tc} 
to study the resonant oscillation of neutrino in the fireball medium. The
combined analysis of both LSND and KARMEN 2 give the oscillation parameters in
the range $0.45\, eV^2 < \delta m^2 < 1\, eV^2$ and $2\times 10^{-3} < \sin^2
2\theta < 7\times 10^{-3}$ with a 90\% confidence level. For our analysis we
consider  $\delta m^2\sim 0.5\, eV^2$ and $\sin^2 2\theta \sim 0.0049$ given
in TABLE\, \ref{lsndtable}.
In this case the $L_e$ and $M_b$ are much higher compared to the one in SNO
and SK. But the resonance length is much smaller than both SNO and SK. For a 10
MeV neutrino propagating in 5 MeV background temperature fireball plasma, we
have $L_e\sim 
4.77 \times 10^{-5}$, $L_{res}\sim 0.7$ km and $M_b\sim 1.33\times 10^{-6}
R^3_7 M_{\odot}$ and for neutrino energy 20 MeV and background temperature 10
MeV, we obtain $L_e\sim  3.28\times 10^{-6}$, $L_{res}\sim 1.4$ km and
$M_b\sim 7.32\times 10^{-7}R^3_7 M_{\odot}$. As the $L_{res}$ is much smaller
compared to the size of 
the fireball, the propagating neutrinos will resonantly oscillate before
coming out of the fireball medium.

\section{conclusions}

We have studied the active-active neutrino oscillation process
$\nu_e\leftrightarrow \nu_{\mu,\tau}$ in the weakly magnetized  $e^-e^+$
plasma of the GRB fireball  
assuming it to be spherical with a radius of 100 to 1000 km
and temperature in the range 3-10 MeV. We further assume that the fireball is
charge neutral due to the presence of protons and their accompanying
electrons. The baryon load of the fireball is solely due to the presence of
almost equal number of protons and neutrons in it. The effective potential for
$\nu_e\leftrightarrow\nu_{\mu,\tau}$ oscillation does not depend on the baryon
content of the fireball, simply because the neutral current contribution to
the neutrino potential is same for $\nu_e$ and $\nu_{\mu,\tau}$. By assuming
charge neutral fireball we have $L_e=L_p$, which we have used to calculate the
baryon content of the fireball. 

We have used the best fit values of the neutrino oscillation parameters from
solar, atmospheric and reactor neutrinos and studied the resonance condition
for the above oscillation process and calculated the lepton asymmetry,
resonance length and the baryon content of the fireball for neutrinos of
energy 5, 10 and 20 MeV and fireball temperature of 3, 5 and 10 MeV. We have
shown that for $\delta m^2$ and $\sin^2 2\theta$ in the solar neutrino (from
SNO) range, the resonance length is large compared to the size of the
fireball which increases with the increase of the neutrino energy and also
the baryon load is low. In this case, probably a few or no resonant oscillation
will take place. But if the $\delta m^2$ and $\sin^2 2\theta$  are in the
atmospheric (SK) or in the reactor neutrino range, there can be many
oscillation before the neutrinos come out of the fireball, so that the average
conversion probability of neutrinos will be $\sim 0.5$. We have also shown
that in these two cases (SK and LSND), the baryon load of the fireball is
neither very low nor very high.  A detail study of the neutrino
 propagation in the GRB fireball is necessary to understand the finer detail
 of the  fireball dynamics. Also this depends on the content of the fireball
 i.e. how much baryon it contains and may affect the dynamics of the jet.

    The GRBs can be detected through GeV or higher energy neutrinos as well as
high energy gamma rays with the present day neutrino and gamma-ray
detectors. All these neutrinos and gammas are produced after the prompt
emission of MeV photons. But these MeV neutrinos due to the collapse of a type
I b,c supernova is similar to the one produced by type II supernova (for
example SN1987A) and are of cosmological distance. These cosmological events
make the MeV neutrino flux very low on earth compared to the one which we had
seen from the supernova SN1987A. Also low energy neutrinos have very low cross
section and combine with the  cosmological distance (low flux) makes the
required detector volume extremely large. So with the present generation
neutrino telescopes it is very difficult to detect these low energy neutrinos.

\vskip2.0cm
{\bf ACKNOWLEDGMENTS} \\
We are thankful to B. Zhang for many useful discussions.
Y.Y.~K and S.~S. thank S. P.~Kim and
APCTP  for a  kind hospitality where this work has been initiated.
The Work of S.~S. is partially supported by  DGAPA-UNAM (Mexico) project
IN101409. Y.Y.K's work is partially supported by KIAS and APCTP in Korea.

\end{document}